# Towards A Sustainable and Ethical Supply Chain Management: The Potential of IoT Solutions

Hardik Sharma, Rajat Garg, Harshini Sewani, Rasha Kashef

Electrical and Computer Engineering
Toronto Metropolitan University
Toronto, Canada
{hardik.sharma, rajat.garg, harshini.sewani, rkashef}@torontomu.ca

*Abstract*—**Globalization has introduced many new challenges making Supply chain management (SCM) complex and huge, for which improvement is needed in many industries. The Internet of Things (IoT) has solved many problems by providing security and traceability with a promising solution for supply chain management. SCM is segregated into different processes, each requiring different types of solutions. IoT devices can solve distributed system problems by creating trustful relationships. Since the whole business industry depends on the trust between different supply chain actors, IoT can provide this trust by making the entire ecosystem much more secure, reliable, and traceable. This paper will discuss how IoT technology has solved problems related to SCM in different areas. Supply chains in different industries, from pharmaceuticals to agriculture supply chain, have different issues and require different solutions. We will discuss problems such as security, tracking, traceability, and warehouse issues. All challenges faced by independent industries regarding the supply chain and how the amalgamation of IoT with other technology will be provided with solutions.**

## I. INTRODUCTION

To maintain a proper business ecosystem from transportation to final payment, it is essential to properly manage the supply chain at each process. The supply chain extensively includes individuals, assets, exercises, and associations engaged with changing crude materials into finished items, fulfilling client orders legitimately and by implication. Numerous variables, including an absence of straightforwardness, disturbances, additional postponements, data contortion, and vulnerabilities, make the complexity of the supply chain even worse. Other challenges that are in achieving a complete supply chain are the capability to handle permeability across start to finish supply chain, trust, atmosphere condition, impractical cultivating rehearses, absence of conspicuousness of resources, fumble of information, vulnerability and in the path for achieving cost efficiency in supply chain effective inventory management act as a primary switch is adjusting demand and supply. IoT has helped solve many problems related to the supply chain. Due to information and communication technology development, IoT technology is increasingly helpful in solving many shortcomings of distributed systems. The data exchange can be done for monitoring, reporting and exchanging between devices. IOT physical devices use sensors like RFID chips and microcontrollers to capture data like location, humidity, speed, etc. The Internet of things has advanced the productivity of creation and dissemination significantly, which made the advancement in big business assist coordination, creation of activity, and exchange. Rapid development in IoT and recent research of SCM have become important research fields. The objective of SCM is to get assembling, storing, and dispersing to the market to upgrade and coordinate general tasks to decrease the cost of the whole chain.

When companies need to confront new partners or merchants or natural disasters, the companies need to face problems related to critical assets and collaboration. Logistic management is a very important factor during global issues on the network, and IoT has proved to be a good solution for making raw bust systems. IoT is one of the trustable solutions for solving many logistic risks. IoT connects various computing gadgets, devices, sensors, or embedded normal-day-life objects with the entire world through the Internet. According to [1], in 2015, there were around 16 billion IoT devices, estimated to increase to 75 billion by 2025, with IoT devices generating nearly 79.4 ZB of data in 2025.

Due to so many benefits and flexibility, IoT frameworks have also become promising solutions to supply chain management systems. Thus, in this paper, we will discuss 1) methodologies and technology that are frequently used in SCM, 2) the supply chain management challenges, 3) most of the problems supply chains face in detail, 4) some of the most important supply chains in various industries, 5) the work completed in different fields and then briefly explain the models used.

## II. PLATFORMS

Common supply chain management and optimization platforms, including blockchain, cloud, and IoT, are discussed next:

### A. Blockchain and IoT

Blockchain technology proved to be a great success when We talk about cryptocurrency, which proved useful in many other fields too. Supply chain management is one of the commonly used applications of blockchain. Blockchain consists of data packets of information referred to as 'Blocks' that have linkage with each other forming a chain [2]. At the ground level, we can say that blockchain is just a chain of blocks. Generally, blocks store three types of information, as given below:

- Information related to time and date
- Information related to participant identification. It is referred to as the digital signature of the participant.
- A unique code, "Hash", is important for identifying each block separately from other blocks.
- Each block contains various headers according to a different type of information; generalizing a block header for supply chain common typically has the following fields:
- Version - to Control the version for a future update.
- Timestamp - for storing the created timestamp of the respective block (when the user starts using the block).
- Merkle root - Hash of all the Transactions/Data.

- Nonce - A randomly generated number to get a hash less than the target hash set by the network.
- A hash of the previous block - the identifier of the previous block in the chain.
- Hash of the current block – the identifier to the current block.

The structure of blocks linked together form a chain, and, in this order, each block stores the Hash of the previous block for reference. Although blockchain can be used for many different supply chains to make it simpler, we will focus on a basic transaction involved in the shipment of products from one place to another, as discussed in [2]. When we combine IoT devices with the blockchain, as a shipment moves from one place to another, it will catch key shipment information discharged from IoT devices. On the other hand, the IoT devices framework would capture the information related to timestamps and current location from the blockchain. The exchange of information between both technologies would be proof of any transaction related to the shipment. Since sensors capture the location of data in the blockchain, it will increase the visibility between different participants, including Original Equipment Manufacturers (OEMs), suppliers, third-party logistic providers, shippers, and warehouses. Suppliers can monitor the shipment and provide additional information related to the consumer, like tracking, location, and date of arrival. The whole lifecycle of a blockchain depends on the linkage with the previous block. However, this thing can make updating information more time-consuming; it provides data integrity, which is very important in the supply chain. This solution will shorten shipment delays, increase the accuracy of predicted shipment time, and strengthen the trust between supplier and consumer. Many research works are conducted using Blockchain with IoT devices for the supply chain; we will discuss its different implementation in the following sections.

### *B. Cloud Computing and IoT*

The data generated by IoT devices can be structured and unstructured and cannot be stored in the traditional Relational Database System. Along with huge storage, it also requires high-performance computation. To achieve such a high-capability system, we have a Cloud-assisted IoT system [3]. We will focus on some basic methods and terms for getting familiar with the Cloud and IoT to create a basic understanding of the system. Cloud Computing integrated with IoT provides efficiency in doing everyday tasks. Cloud Computing is observed as a backend solution for processing huge amounts of data streams and provides computational benefits. This technology will benefit every device connected to a smooth network. The need for cloud computing with IoT is because of the following reasons [4]:

- Sensor Network
- Enable inter-device communication
- Remote processing Provider
- Networking and Communication Protocol.

Cloud computing has three different service models [5]:

- SaaS: SaaS stands for Software as Service, which provides an application that is accessed over the web and is not handled by the company but by the software service provider example, AWS, GCP, or Azure.
- PaaS: PaaS is Platform as a Service which allows users to build and deliver applications without installing IDEs (often expensive).
- IaaS: IaaS is Infrastructure as a service that provides the user with computing capability, network service, virtual private servers, and other resources over the web.

Another type of service model, "DaaS (Data as a Service)," which delivers virtual storage of data and is a special type of IaaS, is also addressed [5]. In [6], Authors have suggested that the IoT architecture is divided into three layers: application, perception, and network. The sensing layer collects information and manages all the sensors like Smart meters, RFID reads, tags, etc. This special type of cloud service is present: Sensing as a Sever. The collected data will be stored over the cloud and provides DaaS on the network layer. This layer can provide computing power, manage all the VMs and ensure protocols are followed. The last layer is an application layer that can provide ease in Supply Chain Management, E-healthcare or VANEts, etc. The applications of Cloud computing with IoT:

- Communication: IoT generates a lot of data in real time. This data generated after processing allows us to take business steps in real time. For this, it requires communication to transfer and process the data. Cloud provides an effective, economical answer to the above-mentioned problem.
- Storage: The data generated from IoT devices can help the business to improve its system. The data can be structured, unstructured, or semi-structured. The traditional database system cannot handle such data. Cloud provides an effective way to store, process, and manage such Big Data.
- Processing Capabilities: IoT devices can collect data but have limited processing capabilities. Cloud services like VMware, prediction algorithms, and data drive decisions can be used to make such complex decisions in real-time without compromising the infrastructure. If the business requires real-time decision-making, cloud computing is the best combination with IoT.
- Scope: The world is quickly moving towards the IoE Internet of Everything. This network brings a lot of devices together, providing heterogeneous information and generating new chances and risks. Cloud-based IoT architecture can provide new emerging services that can deal with real-world scenarios.
- New abilities: The cloud can provide reliability, scalability, interoperability, security, and efficiency. These abilities can be integrated with IoT to deploy low-cost models.

There are many applications related to using the cloud for the supply chain. Since the integration of cloud and IoT devices is possible, many different types of supply chains are using the cloud for better functioning along with the IoT framework

### III. LITERAURE REVIEW

In this section, we will divide the supply chains into different fields and discuss the life cycle, problems, and related research in various industries.

### *A. Pharmaceutical Supply Chain*

The medical industry is one of the most important industries since it is responsible for the health and safety of consumers. We will discuss how IoT devices improve the pharmaceutical supply chain by enhancing traceability and security. Many different factors are important for maintaining a healthy ecosystem in this industry. For some proper medications, it is very important to maintain a certain temperature, such as for vaccines and other clinical or surgical equipment. According to [8], nearly 20% of products are damaged in transportation because they are sensitive to temperature. Similarly, keeping track of the location and providing transparency in the whole chain is also important. Many reports state that fraud related to adulterated medical products in previous years was due to a less secure supply chain [11]. Much research has been done to solve this issue and optimize the pharmaceutical supply chain. The healthcare supply chain has many different processes and nodes. Numerous components, such as lengthy set-up times, asset serious tasks, short timeframe of realistic usability, and high waste creation, make this supply chain different from the other supply chain. The production process takes place in two phases: primary and secondary Levels. Essential creation incorporates the creation of fundamental particles' dynamic segments or drug dynamic fixings. The auxiliary creation likewise incorporates the cycles of change of these dynamic segments and the conveyance to the clients. Depending on different types of equipment, technologies, and system, some challenges need to be addressed by a fully functional pharmaceutical supply chain, such as the crude materials used by suppliers should be properly approved. During the assembly, the whole process should be monitored occasionally, and the transportation should be proper. Since the entire pharmaceutical supply chain depends on trust, it is very important that each node in the chain can keep track of every other node in this industry. One of the problems related to this field is related to fake products. To solve the problem of transparency, traceability in the pharmaceutical supply chain has been enhanced [2]. In this model, they used blockchain and IoT for supply chain management. The main design of the proposed model was dependent on IoT devices for joining the whole chain, and at the backend, cloud computing helps with data security. Each node, from supplier to consumer, needs to authenticate its process; suppliers need to tag the materials utilized in medication.

Blockchain and IoT are important topics in health sector management. Blockchain and IoT for purchase management in the healthcare sector are addressed in [9]; the authors provided a full-edge cloud model for purchase ty, integrity, safety, and traceability [9]. The whole model was divided into different layers, which we will discuss later, and the main part of this was to provide GUI to different nodes for monitoring and tracking process. A GUI on mobile devices is provided in [10]. They used blockchain for optimization, but the focus was to provide a proper platform for each web node. The consumer can track the shipment easily from their android device [10]. The authors used different networking authentication protocols to make the entire system more secure, like login and sharing information according to the user's identification.

Fraud in the medical field is very common and can contaminate the whole chain. Recently related to the current issue of the pandemic, in [11], they explained that global pressure on COVID-19 kits and facemasks had increased the opportunities for counterfeiters. Not only is the health sector supply chain facing this problem, but it has also opened many loopholes to many other sectors. The issue in this problem is related to supply chain security; they worked on increasing the security of IoT devices by introducing a new lightweight security protocol for securing the supply chain. They replicated the attack to show flaws in the previous system and showed their improved protocol based on authenticated encryption.

*B. Agriculture Supply Chain*

Supply chain management in the agriculture industry aims to deliver food efficiently to customers. The importance of SCM in the agriculture sector is as much as necessary as in the medical industry. Farm production is the first block which includes agriculture, stock breeding and fishing; next in the queue is the processing industry which is responsible for maintaining the life of food by reducing pathogenic contaminants. Smart packing and labelling will allow tracking and tracing easy. The logistics and distribution block are responsible for the safe transportation of food by maintaining the food quality considering different environmental characteristics. The next block provides business scope and provides the consumer with the required products. It has been first introduced in Europe with n.178/2002 EU regulation. Through this approach, the product details are managed using barcodes. The traditional process of SCM is very insecure, and the quality of food is not up to the mark. To ensure this complex system does not comprise what consumer gets, it is crucial to have proper management tools and technology. IoT and blockchain have proved to be great technological tools for supply chain management as they have decreased the need for intense manual labour as required in the past. In [13], IoT enabled monitoring system with different sensors is developed for remote areas where the accessibility is very minimum for farmers with good storage facilities to reduce food losses and increase food safety. In [14], to ensure food safety during the entire process of SCM author have aimed to provide a common platform that can be interfaced with third-party application to access all the stakeholders involved in the storage and transportation process. In [15], IoT-based sensing and monitoring are used to ensure fresh fruit supply and less food wastage. In [16], authors have combined the approach of IoT with blockchain technology to build a cloud-based solution for real-time tracking and a trace of material.

*C. Cold Supply Chain*

A cold chain is known as a temperature-controlled supply chain. Maintaining a medical cold chain is challenging because it is risky and complex. In the medical field, maintaining temperature and moisture is one of the essential parts of proper medication. While transporting medication or any other temperature-sensitive device from one place to another, it becomes tougher to monitor the temperature anomalies. Maintaining a cold chain while transporting involves many processes, like modes of transportation, packaging, and equipment for monitoring.

In [8], the cold supply chain challenges are: 1) Risk: knowing the risk at each process, a cold chain is similar to a normal supply chain; every node is dependent on the previous one, and if one connection fails, then it can easily result in loss of quality. Analyzing the risk is necessary and like a pre-requirement of maintaining product quality. So, during the transportation of supply is also important to track which location got compromised or in which process. The time during the anomalies and reaching a stable point in the supply chain is also a risk that needs to be monitored, 2) Complexity: cold supply chains are also very complex and depend on the constant power source for maintaining a constant temperature. When we talk about a real-time supply chain, it will extend from country to country. During that time, maintaining a constant temperature in the whole chain is a very tough and complex task, 3) Governance: the cold chain is divided into fragments involving many different actors for each process. Due to so many parties, there may be a possibility of conflict in objectives, slowing the entire chain, and 4) Product Recall: some medical products or electronics cannot be reused once returned. It is important to maintain each process properly and keep the accuracy of everything according to requirements.

The research in [8] used Wireless Sensor networks and RFID systems. They reduced the number of sensors by combining different sensors into one and named 2G-RFID-Sys for cold chain, allowing the whole system to be more scalable, available, and automatic. In [18], they claimed that although temperature is the leading player across the entire chain, the lag in temperature monitoring is also a major concern for this problem. So, they defined different technologies for better mentoring and divided the hardware design into different modules, which we will discuss in a later section. Big data is a large domain for research and using it with IoT devices can provide a highly integrated framework to the cold supply chain. Authors in [19] designed the smart cold chain system based on Hadoop and Spark. Hadoop and spark are big data technologies that can be used for large data streaming or processing. The whole system can be integrated with many different types of sensors like, from Zigbee to Barcode.

*D. Manufacturing-Industries Supply Chain*

As the demand for automation is increasing in every industry, manufacturing industries also use IoT as the primary technology for making the function fully autonomous. This supply chain is important in terms of the economic perspective of the whole world. The Industrial Internet of Things (IIoT) provides a way for a digital transformation as it has networks of sensors which is used to collect critical data such as production data, and software deployed use this data to turn it into valuable and important insights and reports the efficiency of the manufacturing operation. It also connects instruments and other devices in the same network with industry applications. It is estimated that by 2020, 80% of the Manufacturing supply chain businesses will use cloud-based networks, and this acquisition will increase productivity by 15%, and cost efficiency will also increase by 10% [20]. IoT-enabled devices will provide businesses with real-time insights into the condition of every inventory item and the location and status of supply chain manufacturing units.

IoT has various advantages as it provides a way for cost reduction as it has reduced downtime of machines and optimization of assists and inventory management, which in turn reduces operational costs to the industry and generates more revenue. It enables products to reach the market in a shorter time. Mass customization makes sure that the inventory goes up as there is an increase in the variety of produced stock-keeping units. Not only it helps in increasing the production of products, but it also helps improve safety at the workplace as IoT is paired with wearable devices which constantly monitor workers' health and can prevent injuries. Smart maintenance is a new term introduced by IoT, in which a pre-time warning can be generated for any failure in the manufacturing line, either due to a product or machine. Due to the start of Industrial Revolution 4 (Industry 4.0), the goal is shifted to optimize every aspect of the supply chain. According to Industry 4.0 standards, Industrial Internet of Things (IIoT) communication plotting enormous, grouped arrangements of self-ruling, small, easy and low-force sensor hubs which accumulate and sense information, work together and forward this information to unified cloud backend units for settling preparing and deciding. IoT devices are inactivity touchy and identified with Latency issues, Fog decentralized design was a decent arrangement. In [21], they are utilizing another layer alluded to as the Fog layer was put between IoT devices and gigantic information volume focal cloud centre hubs to handle the difficulties of the monstrous measure of information produced in a rapid clear manner for more circulated and decentralized organization plots as opposed to customary cloud servers. Aside from IoT/Fog decentralized models, other basic issues that are handled are the heterogeneity of hubs and devices of various makers, interoperability from various specialist organizations, the adaptability of control registering and putting away alternatives, and versatility possibilities. IoT/Fog-based decentralized system is additionally appropriate for ongoing machine condition monitoring (MCM), where the pre-handling of crude information is dispersed across haze hubs and arrangement errands likewise take across mist hubs [5]. MCM model embraces a coordinated altering direction method of multipliers (ADMM) calculation as a cross-layer streamlining method for handling idleness, pre-preparing calculation, and prescient support issues. We will talk about its design in another segment.

In [22], through Programmable Logic Controllers (PLC), computerized picture handling to incorporate interlinked and autonomous machines into industry 4.0. Sensor arrangements are likewise accessible however are worked in one use case, and all the current arrangements have been demonstrated to be costly. The freshest methodology is to plug IoT devices into machines, and it empowers brilliant assembling as each item and system is associated with the web and coordinated into an organization. We even use IoT devices to get the information out of straight movements continuously as they are a modest and financially savvy option in contrast to the current arrangements. In [22], they examined the plan and execution of a nonexclusive model for direct movements in an innovation. Gives constant information obtaining. An IoT-device apparatus is planned, which is applied to follow direct movements inside assembling machines and how this methodology has prompted practical methods for machine improvement in contrast with conventional methodologies. Spare part manufacturing industries are also of high importance. Since the spare parts are relatively costly, it is very important to keep the accuracy of each measurement for optimization in this sector by using IoT [23].

IV. DATA SOURCES AND TYPES

On the Internet of Things (IoT), sensors play an integral and vital part as they detect and respond to changes in surroundings [26]. They are used to collect information in various fields like industries, logistics, health and medical, transportation, agriculture, retail, and many more [27]. Different sensors collect different data types that will be used in the IoT framework for making smarter decisions. Some data sources are discussed depending on the different sensor and their usage.

Radio Frequency Identification (RFID) is a wireless microchip device that automatically identifies objects or entities. It uses radio waves to communicate and transmit data between the devices in a network. In many different supply chain models, this data source/module was most used. RFID and IoT are used in many fields, such as agriculture to retail, manufacturing to logistics, and much more. It helps in the real-time management of tagged devices. In the IoT, the RFID module consists of three components:

- RFID tags: RFID tags are transmitters or responders. A tag is a microchip with a coiled antenna used mainly for storing data embedded or attached to an entity that needs to be tracked or managed. Active RFID tags are partially or fully battery-operated, whereas passive RFID tags don't have any power source and work on power provided by the tag reader.
- Reader: it is a transceiver used for activating tags; it is responsible for transmitting data between tags and application software.
- Application system: This component is a data processing system where all the reader and tag activities occur.

The microcontroller is another module used for sensing data and communicating with other IoT-enabled network devices. Arduino is one such microcontroller as it is easy to use, can be used with digital and analog sensors, and can be connected using Wi-Fi or a network-based chip like it. The processing power of Arduino MCU is greater as the processing speed is the main parameter for IoT applications [28]. The power consumption is lower, and it uses a network interface to connect and interact with IoT-enabled devices in a network and send data to the application for necessary processing. Arduino sensors the data in real time and sends it to IoT platforms for processing and storage [29].

Global System for Mobile (GSM) module is a chip that establishes a connection between computing devices. It facilitates a data link in a remote network as it sends data and exchanges IP addresses through SMS messages [30]. GSM module is used with Arduino MCU and is connected to send or receive signals. Together, they act as IoT devices responsible for smart automation. GSM module works with SIM cards where the user sends an SMS from their smartphones to a number registered with the GSM module to convey information [30]. Other IoT-enabled modules include a temperature sensor, mainly used for measuring the amount of heat energy and temperature changes and storing the data collected [31]. The humidity sensor is another IoT-enabled sensor used to measure the amount of water vapour in the environment. Proximity sensors are used for detecting objects (not in contact) near the sensor. They work on the radiation of infrared beams or electromagnetic fields.

V. ANALYTICAL MODELS

In this section, we will discuss the models in much more detail, focusing on the technology, architecture, and implementation.

*A. Pharmaceutical Supply Chain*

In [2], In this model, each node sends the data generated from IoT devices to the server. Using a GPS tracker, a retailer can track the shipment and monitor the production of the assembly line. This model was just for traceability purposes. In [9], the authors provided a proper payment gateway for each process; the system was divided into four layers.

- Data Layer
- IoT Layer
- Blockchain Layer
- Presentation Layer

The first Layer Data Layer in this model is like the input layer. The input is generally related to the purchases between different health sector entities. The system has an engine that creates a table separately for Product, Purchasing, Supplier, and Purchases Detail using the SQL database. It is very important to create tables correctly since the whole architecture is based on them. Next is the IoT layer, which has different types of sensors like temperature, humidity, and location. For experiment purposes in the model [9], all the sensors are controlled by a microcontroller for data transmission to the cloud using Message Queue Telemetry Transport Protocol. Amazon web server and DynamoDB database are used for IoT devices. The third layer in this model is inside the Amazon cloud. For monitoring the purchase process, Hyperledger open-source platform was used to further develop the application and GUI. Hyperledger Composer provides accessibility to create a web application via REST API; this interface is connected with the chain of blocks. So, in the whole system, it is easy for IoT devices to invoke functions implemented in the Blockchain layer. The final layer is like a GUI where different entities of the MQ-135Detect Gasses (to operate at 5V system can log in and do real-time monitoring of the process detect CO). This layer also provides access for creating participants, Fire Sensor, Fire detection used as Infrared order requests, real-time tracking, and smart contracts. In [10], all the sensors must be connected to android devices via Bluetooth. The architecture was divided into three main parts back-end, front-end, and IoT sensors. In the back end, they used the cloud to compile code on a virtual machine. They used the PostgreSQL database; for communication between the front end, they used JSON parsing. The front end communicates with the server using REST API. A client using android devices can easily overwrite the sensors in this model depending on the permissions [10]. The reading of different sensors was made visible to both the receiver and sender to prove transparency in the system.

*B. Agriculture Supply Chain models*

In [13], the authors have discussed the importance of agriculture, one of the main income sectors in developing countries. In this, they have developed an IoT-based system that will use a different sensor to collect data and send out the inside location of the warehouse and environment details to farmers via SMS or Email. The IoT nodes are developed for the experiment using the ESP32 Wi-Fi module and sensors [13]. These sensors are interfaced with a WiFi module and installed in the warehouse. The software implementation includes ESP32 Programming, which uses the Arduino IDE platform and Node-Red Dashboard, which helps collect and visualize data from the sensor. The first step is to preheat the sensor for 5 seconds and establish a Wi-Fi connection. The next step is to establish an MQTT protocol (Mosquito Broker) which is between publishers (Sensor) and subscribers (the device that has been subscribed). After that, sensors will calculate the temperature, humidity, and CO value and detect fire, motion, and shock. These data will be published on the MQTT broker. These data will be sent out to Node RFD local IP, which will display a dashboard showing a live demo of sensor details to farmers. These details are even mailed after every 30 minutes. With the help of this experiment, farmers can preserve grains and vegetables. This will reduce food wastage and increase food safety. In [14], authors have monitored the environmental conditions where the food is stored using different sensors. The value obtained from the sensors is first compared with the threshold value before sending it to the remote system. All the plotting of data and graph creation is done at the remote system. GUI is designed to provide better user monitoring. This will help the end-user, who can be anybody that handles the shipment of the food product, to get the data values, and if any emergency occurs, then it can take necessary steps. The authors have created different interfaces in the GUI. To secure the system, they have created a login page and A different page for monitoring sensor values. Log Generation using Bluetooth as well as Wi-Fi. For future scope, they want to build an app capable of processing the data received from the sensor and sending out alerts for any damage to the food. This heterogeneous sensor will ensure food safety at any stage of the supply chain.

In [15], the authors deal with problems related to transportation and the distribution of fresh fruit. IoT-based sensors and monitoring tools are used during the transportation process. Current transportation and distribution of food spoilage and waste is around 12%. The existing system suffers from problems like information privacy needs are challenges; along with that, it is time-consuming and no continuous monitoring using magnetic induction and scan coil detector. To overcome the issues, the authors have used an MQ3 sensor and IoT devices (Arduino IDE). The paper is divided into three sections: a) Sensing and output display module where the MQ3 sensor will send out information to the Arduino controller about fruits condition. If the fruit is spoiled, the message is on the LCD monitor, and the buzzer will start ringing. This GSM module will send an alert to the driver with the package number so the robot sensor will pick up and dispose of the damaged food.

In [16], researchers with an amalgamation of RFID, IoT, and blockchain have proposed a cloud-based portal for real-time tracking and tracing of logistics and supply chains. Existing tracking systems like GPS, GTN, and RFID cannot track such scenarios as recognizing the products within a container that are opened or lost. To tackle such a situation, IoT, along with the Industrial Internet platform and trackers, will provide a smooth production process and management. This paper is based on a case study approach where the authors have explored the supply chain network within an energy-based company for a project in Finland. The research is aimed to do the following task i) Identification of functional and non-functional requirements of logistics in project business ii) Develop a portal by combining IoT, RFID, and blockchain for tracking and trace of products in project business. The blockchain architecture is developed using Ethereum, an open-source platform for blockchain applications. Ethereum consists of blocks that are connected to a blockchain. This architecture uses "Ether," which is a token handling the task of payment for work completed, like creating and verifying blocks. The Dapp is a decentralized app that will help deploy the application in the front and back end. The smart contract solidity in the Ethereum deployment block is responsible for creating computerized transaction agreements between parties to ensure supply chain partnership. To know about transactions, the authors have developed an app and web user interface where one needs to fill in the basic logistic-related transaction information like GPS location of the device, generated label ID, user id, and time stamp. Blockchain generation is done within the Ethereum network using the above information. Truck IoT, tracking devices, and AIS. The cloud portal developed by the authors has data related to customers, purchase orders, projects, shipments, and deliveries, which are easily handled at a centralized level. This will solve the issue of verification with a fair response time. They used Kovan (test net for the Ethereum application) to test the transaction. For tracking, the authors have used present in the blockchain, which assures transparency and accurate traceability. Their architecture contains three important components i) API, ii) Controller and iii) Blockchain. This layered architecture will provide a complete history of the supply chain from farm to fork. The only pre-condition is that all the components involved in the supply chain are registered under a blockchain and have access to key pairs. Table 2 shows the performance of AgriBlockIoT in terms of Latency, Network traffic, and CPU load.

*C. Cold Supply Chain models*

The model used in [8], WSN, along with RFID, keeps a record of fluctuation in temperature, thoroughly monitors everything, and triggers pre-alerts in case of failures. This model aims at maintaining a direct ecosystem between nodes. The model identifies each unit separately and uses thermal packing for a temperature-controlled environment. The sensor networks keep a note of the temperature of units at proper intervals. Their model takes care of different problems, from risk to governance, in achieving a healthy cold supply chain through the Temperature monitoring system. The finished products are sent for quality analysis and then to the consumers. In [18], the problem of temperature monitoring lag was solved by dividing the whole system into four main modules:

- Microprocessor module: The STM32L1 series controller is used. For less power consumption, the module controls the functioning of the whole system
- Position module: The L76C series GNSS module was used; it was connected to an accelerometer to get the unit's perfect position.
- Sensor module: Different sensors from temperature and humidity are integrated with this module to make the whole system intelligent.
- Communication module: This module was used to transmit data at high speed and with lower power consumption.

In [19], the authors divided the data sources into two types: application-based and external. External data source values are passed through the import manager, while the sensor data readings are passed to the abstraction layer and then to spark streaming layers. The data storage manager sends the filtered data to the HDFS system. Since visualization is important for getting insights from the model and making predictions related to future anomalies, both type of processed data is sent to visualization tools.

### D. Manufacturing-Industries Supply Chain models

The model explained in [22] is related to the automation as discussed in [21], is an IoT/Fog/Cloud MCM system model where the MCM layout consists of M autonomous machine parts or machines and along with this, there is an IoT gateway for each of them [21]. This model consists of three different layers, i.e., IoT layer, Fog layer, and Decision or cloud layer. Each layer function is defined as follows:

1) IoT Layer: From each machine, this layer collects raw data measurements that are performed by N (N>0) sensor hubs (which appeared as specks in the figure) and sends them to their passage hubs. The gathered crude information is identified with machine data (type, includes, the reason for use, and so on). In this way, each machine has N hubs mounted over it. They have restricted computational abilities and energy holds, and the hubs discuss just with their door hubs in remote mode (e.g., Wi-Fi availability).

2) Fog Layer: In this layer, M gateway nodes (fog nodes or workers) are mounted on the highest point of each machine. These hubs generally perform consolidating, figuring, and accumulating assignments, delivering order information results, and sending them to ace hubs (cloud or center workers). These hubs are all the more remarkable when contrasted with the sensors. These cross-layer computational exchanges can be performed on account of the chance of every door to get information directions from the cloud worker. The information result likewise described as functions which are prepared additionally reveals to us machine use condition (power utilization, breakdown log reports, and so on). The offbeat altering direction method of multipliers (ADMM) calculation is answerable for cross-layer conveyed streamlining for the information trade guidelines [21].

3) Decision Layer: This layer is liable for dynamic, requesting computational assignments, and applying any recommended improvement method to decide the correct machine choice outcome, identified with the machine's ability and worldwide market dependability. This layer likewise comprises a cloud worker (Master hub), which gives basic settled data to practical flexible chain exercises—collected characterization information from implementation of state distinguishing proof in straight movement from measures in genuine creation conditions which depends on a coordinating model.

Machine measures are performed by straight hub on vertical and flat vehicle frameworks on which we have set our core interest. The theoretical model relies on a line portion, which all in all characterized as a piece of a line that is limited by two unique endpoints [22] and is commonly named as follows: when there are two endpoints to be specific, An and B then the line fragment will be composed as AB or BA which altogether relies upon the direction of the section. The straight development inside a cycle of a machine characterizes the direction of the fragment. When the underlying point A and state B is the endpoint of the development, then the development from A to B is section AB. A similar development back from B to the underlying point is named the fragment BA—this grouping of state-development rehashes in a perfect world endless occasion. In our model, the separation is a fixed boundary, as we probably know from the meaning of a fragment that separations AB and BA are equivalent. The span of states An or B and the length of developments AB or BA are not fixed. Along these lines, the speed of developments and holding up times in states can fluctuate as per the cycle. In [23], the model was designed in such a way that Many businesses can benefit from using IoT in the extra parts flexibly chain because the basic advancements need no discoveries, which helps the Original Equipment Manufacturers (OEM) and end-clients to accomplish higher edges [13]. Save parts and hardware fix administrations. Gear OEMs can distantly screen resource well-being in their clients' plants, envision disappointment, execute fixes before the disappointment happens, and order the parts can be monitored by IoT-enabled conditions. It is Integration of big data analytics with IoT tools and spare parts predictions is more feasible as demand forecasting is highly unpredictable. After the item is forecasted, an application of reordering policy is issued and sent to the OEM in conventional demand forecasting. After that, the demanded spare part is dispatched by OEM, which takes considerable time. Here IoT platforms can be used, and OEMs will have real-time data about the parts, and in turn, processes like policy issues and lag time can be eliminated. It is also beneficial for monitoring asset health for the prediction of failure and repairs, which reduces maintenance costs and usage of inventory as the OEMs have real-time status about the equipment they need.

## VI. DISCUSSION

IoT is quite a trending technology and has provided improvement across industries [33]-[37]. IoT can be anything, such as artificial biochip transponders in farm animals or automobiles with sensors. IoT provides a lot of ease in our daily life; with it, any work can be easily done at just our fingertips. Therefore, it is very convenient, and these benefits will keep attracting people to IoT technology. During the past few decades, large companies have been obsessed with collecting and processing data about their customer and internal processes [38][39]. This is because businesses are trying to involve in analysis to gain a competitive edge. This data can be collected through continuous monitoring and analysis using machine learning technologies [40][41].

It was difficult for industries to keep track of suppliers and modes of transport. Without IoT, we can achieve real-time visibility at each step of SCM. Various papers discussed above have improved product quality by maintaining temperature, humidity, and transportation of products to the consumer. It will be easier for businesses to provide loss protection for lost products with IoT. Everything comes with drawbacks; small companies cannot expand and automate their operations like Amazon. There are, of course, cheaper solutions present that small industries can adapt to. Apart from this, IoT has been booming technology for the gold mining industry, wherein it is important to maintain the temperature of any product. Traditional cold chains require the physical presence of a person to monitor the temperature. With IoT, managers can analyze and toggle temperature settings as per requirement, reacting to changes in local climate, damage to the packaging, unpredicted delays, and human error. Even in the especially rare scenarios where temperature errors are recognized but cannot be addressed, the level of detail provided by IoT technologies enables managers to react to potential concerns sooner before they become a major crisis. Everything comes with a drawback. By adapting to IoT technology, we will relinquish the quality control process by trusting sensors and machines to identify the machine's needs and decide. Traditionally decisions are made by people by observing the situation. That is a big change in how business has been done in the past. Another issue is that the data obtained from the sensor may be incomplete and insufficient. Consider an example, data obtained through offline processes and partners who are not running fully digital operations. The data from these sources play a major role in any business process. Security is a major concern for any industry if there is any cyber-attack [42]-[46]; in that case, the industry will stand still and can cause huge loss [47][48]. SCM is a major part of every consumer-based industry; bringing revolution to its management will indeed increase the competitive edge of any enterprise. IoT is one solution that can far-reaching improvement in SCM by providing numerous benefits, bolstering any company's business.

VII. CONCLUSION AND FUTURE DIRECTIONS

The supply chain management industry is facing numerous challenges in today's globalized world. However, the Internet of Things (IoT) has emerged as a promising solution to address many of these challenges by providing security, traceability, and reliable relationships between supply chain actors. By leveraging IoT devices, businesses can create a more secure and trustworthy ecosystem that addresses the unique needs of different industries, including pharmaceuticals, agriculture, and warehouse management. This paper has demonstrated the potential of IoT technology to improve supply chain management processes, highlighting the importance of collaboration between different industries to achieve more efficient and effective supply chain management. As such, we can conclude that IoT technology is essential for the future of supply chain management, and its continued development and adoption will be critical for businesses seeking to remain competitive in today's ever-changing marketplace.

There is a growing need for the integration of IoT with other emerging technologies such as blockchain, artificial intelligence, and big data analytics. By leveraging these technologies in conjunction with IoT, businesses can achieve even greater levels of security, transparency, and efficiency in their supply chain operations. In addition, with growing focus on sustainability and ethical practices in supply chain management, IoT can play a key role in ensuring compliance with regulations and ethical standards by providing real-time tracking and monitoring of products and materials, as well as enabling more efficient and sustainable use of resources. As the global supply chain becomes increasingly complex and interconnected, there is a need for greater collaboration and standardization across industries and regions. IoT can help facilitate this by providing a common platform for communication, data sharing, and collaboration between different supply chain actors. The future of IoT in supply chain management is promising, and continued innovation and collaboration across industries will be essential for achieving more efficient, sustainable, and secure supply chain operations.